\documentclass[sigplan,nonacm]{acmart}
\usepackage{tikz}
\usepackage{tcolorbox}
\usepackage{tabularx}
\usepackage{graphicx}
\usepackage{xspace}
\usepackage{xcolor}
\usepackage{subcaption}
\usepackage{algorithm}
\usepackage{algorithmic}
\usepackage{tikz}
\usetikzlibrary{shapes.geometric}
\usetikzlibrary{shapes.misc}
\usetikzlibrary{decorations.pathreplacing}

\definecolor{darkgreen}{RGB}{30, 120, 30}
\newcommand{\name}{{xPU-Shark}\xspace}
\newcommand{\fullname}{xPU-Shark\xspace}

\newcommand{\uarch}{{microarchitecture}\xspace}
\newcommand{\uarchal}{{microarchitectural}\xspace}
\newcommand{\Uarchal}{{Microarchitectural}\xspace}
\newcommand{\assembly}{{machine-code}\xspace}
\newcommand{\datacache}{{VMEM}\xspace}
\newcommand{\Datacache}{{VMEM}\xspace}
\newcommand{\MXU}{{MXU}\xspace}

\newcommand{\modelone}{LLM-Small\xspace}
\newcommand{\modeltwo}{LLM-Medium\xspace}
\newcommand{\modelthree}{LLM-Big\xspace}

\newcommand{\qXHLO}{\fbox{Q1}\xspace}
\newcommand{\qMXU}{\fbox{Q2}\xspace}
\newcommand{\qMEM}{\fbox{Q3}\xspace}
\newcommand{\qACT}{\fbox{Q4}\xspace}

\newcommand{\cSW}{\fbox{C1}\xspace}
\newcommand{\cNOCOMP}{\fbox{C2}\xspace}

\newcommand{\dmaissue}{ISSUE\xspace}
\newcommand{\dmawait}{WAIT\xspace}

\newcommand{\AllGather}{All-Gather\xspace}

\def\ie{{i.e.\ }}
\def\eg{{e.g.,\ }}

\def\etc{etc.\xspace}

\newif\ifdraft
\draftfalse

\newcommand{\asaf}[1]{\ifdraft{\noindent{\textcolor{red}{\bf \fbox{Asaf} {\it#1}}}}\fi}

\newcommand{\yannis}[1]{\ifdraft{\noindent{\textcolor{orange}{\bf \fbox{YZ} {\it#1}}}}\fi}
\newcommand{\amanda}[1]{\ifdraft{\noindent{\textcolor{purple}{\bf \fbox{AT} {\it#1}}}}\fi}

\usepackage{enumitem}
\setitemize{noitemsep,topsep=0pt,parsep=0pt,partopsep=0pt}
\setlistdepth{5}
\setenumerate{noitemsep,topsep=0pt,parsep=0pt,partopsep=0pt}
\setdescription{itemsep=0pt,topsep=0pt,parsep=0pt,partopsep=0pt,itemindent=0em,leftmargin=0.5em}

\newcommand{\draftfigure}[1]{%
  \begin{tikzpicture}
    \node[anchor=south west] (image) at (0,0) {#1};
    \ifdraft\node[draw, red,ultra thick,rounded corners,anchor=center, scale=3, opacity=0.25, rotate=45] at (image.center) {DRAFT};\fi
    \end{tikzpicture}
}

\begin{document}

\title{Fake Runs, Real Fixes – Analyzing xPU Performance Through Simulation}

\author{Ioannis Zarkadas}
\authornote{Authors contributed equally, order decided by coin-toss.}
\authornote{Work done while at Google.}
\affiliation{%
  \institution{Columbia University}
  \country{}
}

\author{Amanda Tomlinson}
\authornotemark[1]
\authornotemark[2]
\affiliation{%
  \institution{University of California, San Diego}
  \country{}
}

\author{Asaf Cidon}
\affiliation{%
  \institution{Columbia University}
  \country{}
}

\author{Baris Kasikci}
\affiliation{%
  \institution{University of Washington}
  \country{}
}

\author{Ofir Weisse}
\affiliation{%
  \institution{Google}
  \country{}
}

\begin{abstract}

As models become larger, ML accelerators are a scarce resource whose performance must be continually optimized to improve efficiency. Existing performance analysis tools are coarse grained, and fail to capture model performance at the \assembly level. In addition, these tools often do not provide specific recommendations for optimizations. 
We present \name, a fine-grained methodology for analyzing ML models at the \assembly level that provides actionable optimization suggestions.  
Our core insight is to use a hardware-level simulator, an artifact of the hardware design process that we can re-purpose for performance analysis. \name captures traces from production deployments running on accelerators and replays them in a modified \uarch simulator to gain low-level insights into the model's performance.
We implement \name for our in-house accelerator and used it to analyze the performance of several of our production LLMs, revealing several previously-unknown \uarch inefficiencies. Leveraging these insights, we optimize a common communication collective by up to 15\% and reduce token generation latency by up to 4.1\%.

\end{abstract}

\maketitle

\section{Introduction}

As generative artificial intelligence (AI), is projected to become a trillion dollar market by 2032~\cite{genai_bloomberg_study}, an increasing number of companies invest in developing ML accelerators.
In addition to the established chip designers, such as NVIDIA~\cite{nvidia}, AMD~\cite{amd} and Intel~\cite{intel}, hyperscalers like Google~\cite{tpu, tpu_v4}, Meta~\cite{meta-mtia}, and Amazon~\cite{aws_inferentia, aws_trainium}, startups like Cerebras~\cite{cerebras_wse3}, and academic researchers~\cite{mit_accel_eyeriss, harvard_accel_minerva, stanford_accel_eie} have also developed custom ML chips.

A complex ecosystem of tools have been built around these accelerators to support ML development.
High-level frameworks like TensorFlow~\cite{tensorflow}, JAX~\cite{jax}, and PyTorch~\cite{pytorch} allow engineers to express machine learning models with simple APIs. 
These high-level models are then translated into ML intermediate representations like MLIR~\cite{mlir} or OpenXLA StableHLO~\cite{xla}.
These mid-level representations provide a layer of indirection between high-level ML frameworks and machine-level code, allowing compilers to target custom hardware from a reduced set of intermediate representations.
These portable mid-level representations are then compiled into the byte-code which runs on the ML accelerator. 
The development of each of these levels of abstraction requires a huge engineering effort, and inefficiencies introduced at any level can cause performance degradation for the model.
The companies that offer generative AI services are often doing so at a massive scale (for example, the infrastructure to provide inference for Microsoft's Bing AI chatbot is estimated to cost \$4 billion~\cite{genai-cost}), meaning that even a small degradation in performance can lead to large capital losses. Some companies such as DeepSeek are even implementing certain features directly in \assembly~\cite{deepseek_uses_ptx}, showcasing the importance of low-level optimizations.

Because of the utmost importance of model performance, ML engineers need robust profiling and optimization tools in
order to squeeze the maximum performance out of  accelerator hardware.
However, existing performance profiling tools fall short in analyzing low-level performance. 
Many tools (\eg Nvidia NSight Systems~\cite{nsight_systems}, Tensorboard~\cite{tensorboard}) provide coarse-grained metrics on the high-level operations (HLOs) of the intermediate representation by leveraging the accelerator's performance monitoring unit (PMU). 
This is useful for revealing inefficient HLOs, but offers little visibility into the interaction between the code and hardware. Moreover, the granularity of metrics provided by the PMU is constrained by its buffer size.

Other tools focus on finding places in the code that cause hardware stalls by sampling program counters (\eg Nvidia CUPTI~\cite{vtune}, Intel VTune~\cite{vtune}).  However, the root cause for a stall can be hard to connect to a specific stalling instruction, and thus knowing where the stalls happen offers little insight on how to fix them. 
In addition, stall PC sampling requires hardware support which is not always available.
Another category of tools (\eg NVBit~\cite{nvbit}, CUDAAdvisor~\cite{cuda_advisor}, ValueExpert~\cite{valueexpert}) are based on binary instrumentation, which records information about every low-level instruction executed in the accelerator. 
While this information enables fine-grained analysis of the software, it changes the hardware utilization characteristics at runtime. 
Finally, instrumentation typically requires re-compilation, which makes it hard to apply in a production setting where models and code are constantly changing.

\begin{figure}[h]
    \centering
    \includegraphics[width=0.44\textwidth]{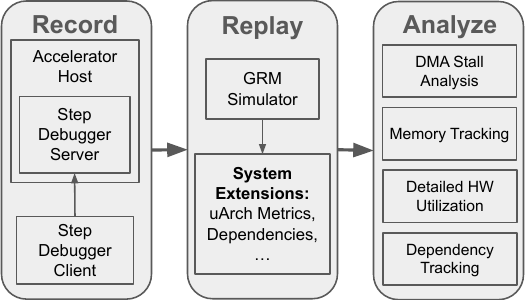}
    \caption{Overview of \name.}
    \label{fig:system-arch}

\end{figure}

\amanda{our main point of novelty: the reuse of the GRM to enable record \& replay in accelerators.}
To fill this gap, we present \fullname~(\name), a novel methodology to analyze the \uarchal efficiency of ML accelerators in the context of a hyperscalar datacenter.
As shown in Figure~\ref{fig:system-arch}, \name enables a ``record and replay'' style of profiling by recognizing that a common artifact of the accelerator design process, a Golden Reference Model (GRM), can be repurposed as an Instruction Set Architecture (ISA) level simulator. \asaf{I wonder if early on in the paper it would make sense to address the "elephant in the room" here -- which is that only the designer of the accelerator has access to the GRM. Maybe we should explicitly say somewhere in the intro that given the economics of AI more and more cloud operators are developing their own silicon, and therefore these techniques would be relevant for such players. In addition, perhaps our work might motivate some chip designers to release their GRM to their users so they can use it to better analyze their models' performance}
This methodology allows the capture of fine-grained details for both the software (\eg instruction dependencies, memory accesses) and the hardware (\eg compute units utilization, memory stalls).
\name first uses a step debugger to capture traces from production deployments of our in-house ML models, then replays them in a modified ISA-level simulator.
We then use the data produced from the ISA-simulator to build a series of performance analyses, and automatically suggest performance optimizations. \asaf{maybe provide a bit more detail here about how you can generate these optimizations automatically?}

In this paper, we describe how we used \name to optimize our large language models (LLMs).
First, we describe how \name can help identify and visualize inefficient memory transfers (DMAs).
Second, we use \name to analyze the utilization and fragmentation of our accelerator's \datacache and provide fine-grained instruction-by-instruction utilization information for various compute units of our accelerators, enabling our engineers to reason about how their code performs at the \assembly level. %
Third, we use \name to analyze the dependencies of each instruction to automatically suggest optimizations via instruction reordering. \asaf{this is very vague}
Although our in-house LLMs are already highly optimized, \name revealed several previously unknown inefficiencies that amounted to up to 4.1\% of token generation time.

In summary, we make the following contributions:
\begin{itemize}
    \item We introduce \name, a new performance analysis framework that leverages \uarchal simulation to provide performance insights. To the best of our knowledge, \name is the first work that extends a \uarch simulator to tune performance on a \emph{specific} accelerator architecture.
    \item We implement \name for our in-house accelerator and construct several performance analyses based on the data we collect with it, which are very hard or impossible to implement with existing tools. \name revealed several inefficiencies in our LLMs, even though they have been thoroughly optimized already.
    \item Using \name's suggestions, we optimize a common communication collective (All-Gather) by 15\% \asaf{why is the 15\% result not emphasized in the abstract..? 15\% >> 4\%} and decrease the generation time of our LLMs by up to 4.1\%. These LLMs are deployed at huge scale internally and each percent improvement in model serving leads to significant savings in total cost.
    
\end{itemize}

\section{Background and Related Work}

\begin{figure*}[ht]
    \centering
    \begin{subfigure}[c]{0.49\textwidth}
    \includegraphics[width=\textwidth]{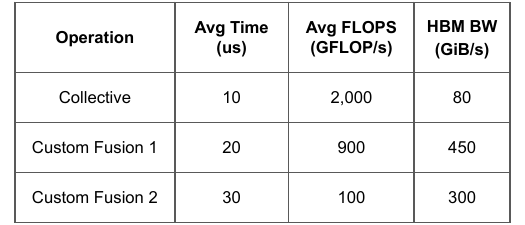}
    \caption{Coarse Grained}
    \label{fig:motivation:coarse-grained}
    \end{subfigure}
    \begin{subfigure}[c]{0.49\textwidth}
    \includegraphics[width=\textwidth]{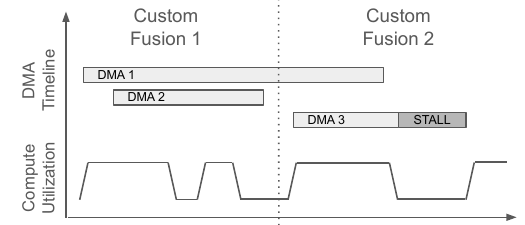}
    \caption{Fine Grained}
    \label{fig:motivation:fine-grained}
    \end{subfigure}
    \caption{A toy example showing a coarse-grained vs fine-grained analysis. Many existing tools provide coarse-grained information at the kernel or HLO level, while deep optimization requires a fine-grained view. }
    \label{fig:motivation}
\end{figure*}

In this section, we provide a brief overview of the ML software development landscape (\S\ref{background:ml-ecosystem}). We then discuss the growing need for optimizing accelerator performance and how profilers are a crucial tool in doing so (\S\ref{background:profilers}).

\subsection{ML Software Stack}
\label{background:ml-ecosystem}

High-level frameworks like TensorFlow~\cite{tensorflow}, JAX~\cite{jax} and PyTorch~\cite{pytorch} provide simple APIs for constructing ML models.%
High-level models are then translated into intermediate representations like MLIR~\cite{mlir} and OpenXLA StableHLO~\cite{xla-hlo}. 
The intermediate representation is usually a computation graph consisting of higher level operations (HLOs), like ``matmul'' and ``transpose''. 
These mid-level representations provide a layer of indirection between the ML frameworks and ML compilers~\cite{xla, openxla, iree, tvm}.
The compiler then uses these representations to generate accelerator code (\eg CUDA kernels, PTX, SASS) in a process known as \emph{lowering}.

Given the very high costs of deploying and operating AI infrastructure, there is a lot of work on squeezing more performance out of existing accelerators. Performance experts focus on constructing faster kernels through careful orchestration of the hardware and especially memory, via techniques like FlashAttention and PagedAttention~\cite{flash_attention, flash_attention_2, flash_attention_3, paged_attention}. Compiler engineers invent new frameworks that expose greater control of the hardware, like Triton~\cite{triton} and Pallas~\cite{jax_pallas}. And at a cluster level, resource managers like Pathways~\cite{pathways} take ML workload characteristics into account when making global scheduling decisions.

\subsection{Profilers}
\label{background:profilers}

To allow developers to debug and optimize their workloads, vendors and academics have developed a number of performance profiling tools.

\textbf{Motivating Example.}
In Figure~\ref{fig:motivation} we show a simplified example of an inefficiency which would be hard to detect with current profiling tools.
Existing tools present performance information in a coarse-grained format, usually listing an operation (the HLO or kernel name, like ``Custom Fusion 1") and any aggregated statistics of interest (like runtime, FLOPS, or utilized memory bandwidth).
These tools show which operations have low performance, but do not provide the reasons for low performance or clues on how to increase performance.
In contrast, a low-level optimization tool like \name can offer deeper insight into each \uarchal component of the accelerator.
In our example, the course grained analysis shows that the second custom fusion HLO has lower than expected average FLOPS. 
We may also be able to see that hardware bandwidth utilization can be improved which gives us some clue that the memory subsystem is responsible for the low performance.
The fine grained analysis gives the information we need to fully analyze and fix the issue with our custom HLO.
In Figure~\ref{fig:motivation:fine-grained} we can see each memory transfer along with detailed compute utilization metrics. \asaf{It's not clear why custom fusion 2 is executed after custom fusion 1? Maybe provide a bit more narrative for what each one of the functions might be doing? I would recommend making this example more concrete, with some real functions, and explain what they're actually trying to do, why are they DMAing, what are they actually computing. The more concrete you can be the better, right now it's too vague.}
We can see that the compute utilization is low because the hardware is stalling when issuing DMA 3.
We can also see that to fix the issue, we need to issue the DMA earlier, preferably during or before the first custom fusion HLO (cross-HLO optimization). 
Finding opportunities for optimization across HLO's is especially difficult with coarse grained tools, as it is not clear how different operations interact from aggregated statistics. 
Here we showed a simple example of how a fine grained view into the \uarchal utilization of the accelerator can make performance analysis much easier, and we now give an overview of existing profilers and their capabilities.

\textbf{PMU.}
Vendors typically provide hardware support to record performance events, the performance monitoring unit (PMU), along with profilers that leverage it.
The PMU has a number of registers (performance counters) which can record performance events such as cache hits, or collect statistics like cycles executed within a function. 
Profilers that use the PMU include Nsight Systems~\cite{nsight_compute}, nvprof~\cite{nvprof}, Intel VTune~\cite{vtune}, AMD ROC-profiler~\cite{rocprofiler} and Google Tensorboard~\cite{tensorboard}.
More specifically, Nsight offers utilization metrics at the level of a CUDA kernel and Tensorboard at the level of a HLO. 
These tools are usually lightweight and rely on the PMU to provide performance metrics, which is constrained both in granularity, because of its limited buffer size, and in variety, as it typically allows recording a limited number of performance counters simultaneously~\cite{nvidia_multi_pass}.

\textbf{Program counter sampling.} Besides the PMU, some vendors (\eg Nvidia) provide program counter (PC) sampling and stall attribution. 
PC sampling requires additional hardware support to sample the program counter during execution.
If the code is stalling at the time of the sample, the hardware may also provide the reason for the stall (\eg memory stall, synchronization stall). 
Tools like CUPTI~\cite{cupti}, VTune~\cite{vtune}, HPCToolkit~\cite{hpctoolkit} and DrGPU~\cite{drgpu} use PC sampling to collect stack traces and their stall reasons, coalesce them and suggest optimization strategies. 
While this approach can be useful for finding opportunities to improve performance, it has a number of disadvantages. 
First, it requires additional hardware to sample running code and attribute stall reasons correctly, which is not available in all accelerators, including ours. 
Second, while PC sampling can pinpoint various inefficiencies like memory transfer stalls, it does not show the root cause instruction. For instance, a stalled instruction waiting for a memory transfer is the symptom, not the cause.
Finally, it does not provide any information on how well the hardware is utilized, as it focuses on stalls and not on hardware utilization.

\paragraph{Instrumentation.} Another category of tools use binary instrumentation to gain performance insights on a more microscopic level, albeit with great overhead. 
Binary instrumentation engines such as Nvidia NVBit~\cite{nvbit}, SASSI~\cite{sassi}, Sanitizer API~\cite{sanitizer_api}, Intel GTPin~\cite{gtpin} and LLVM~\cite{llvm} can change code at a very low level, so that every instruction can be recorded. 
This approach is used to analyze the behavior of the software at very high detail. 
For example, VALUEEXPECT~\cite{valueexpert} traces every load and store instruction to discover inefficient patterns in the data, like hidden sparsity or repeated computation. 
CUDAAdvisor~\cite{cuda_advisor} traces memory accesses to reveal memory metrics such as reuse distance. 
While these approaches are useful to explore the inner workings of a program running on an accelerator, they offer no insight on how well the program is using the underlying hardware, as instrumenting the program totally changes its execution characteristics which renders the PMU useless. 
In addition, instrumentation typically requires re-compilation, which makes it hard to use in the context of a hyperscalar where compilation is complex, while code and dependencies change frequently.

\textbf{Cross-cutting.} Last but not least, some tools combine multiple techniques to provide a more detailed performance analysis. GPA~\cite{gpa} combines PC sampling with instrumentation to detect inefficient parts of the code, analyze their dependencies and suggest root causes. It is a powerful optimization tool but does not provide specific optimization suggestions (\ie move this instruction here). In addition, it relies on hardware support for PC sampling, which is not available in every accelerator, including ours. NVidia NSight Compute~\cite{nsight_compute} is the most comprehensive tool we know, analyzing CUDA kernels down to the PTX level, detecting instruction dependencies, and warning of inefficiencies. However, it analyzes only one kernel at a time, misses Nvidia SASS-level insights, and remains opaque due to its proprietary nature.

\section{Design Requirements}
\label{sec:design_requirements}

To get the best performance out of an accelerator, it is important to have good visibility of how the software interacts with the hardware, at a fine granularity. Based on these observations, we pose the design requirements for \name as a series of questions that an ideal low-level profiler should answer.

\paragraph{\qXHLO Are there optimization opportunities inside and across HLO boundaries? }
Higher level performance analysis tools only present aggregate metrics for each HLO, 
even though data dependencies often span HLO boundaries. Can a detailed analysis below this abstraction unlock new opportunities?

\paragraph{\qMXU What is the instantaneous utilization of individual \uarchal units?}
ML workloads require heavy matrix and vector multiplications and manipulations which need to be carefully orchestrated to achieve optimal use of the accelerator. 
Aggregated statistics can hide opportunities to fully utilize the accelerator hardware.

\paragraph{\qMEM Is our \datacache properly utilized to alleviate memory bottlenecks?} Data-caches help alleviate the memory bottleneck, and bridge the gap between relatively slow memory and incredibly fast compute. %
Efficient use of the \datacache is perhaps the most crucial element of achieving maximum performance.

\paragraph{\qACT Can the tool provide actionable insights?}
Performance analyses and visualizations can help users better understand how the \assembly interacts with the hardware and spot inefficiencies. 
Existing tools can often point to general causes, but cannot suggest specific actions to fix them. \newline

In addition to these questions, we face several challenges in the context of a hyperscaler:

\paragraph{\cSW Use software, not hardware.}
Techniques like PC sampling help in pinpointing inefficiencies, but they require hardware support that is not available in all accelerators, including ours. 
We want the solution to be implementable entirely in software, so that it can be applied immediately to our entire accelerator fleet.

\paragraph{\cNOCOMP Avoid recompilation.} Many tools that analyze performance at the lowest level commonly need model code to be compiled with special flags that instrument the resulting program.
However, large ML models often have complex compilation pipelines that are slow and hard to modify. 
In addition, code and dependencies are constantly changing, making it harder to accurately recompile production models.

\section{Design and Implementation}
\label{sec:design}

\name consists of an \textit{execution recorder}, a \textit{replayer}, and an \textit{analyzer} (Figure~ \ref{fig:system-arch}). The execution recorder (\S\ref{subsec:design_recorder}) uses the hardware step debugger to break in the middle of the model execution and record traces of the machine-instructions executing the ML model. The traces include the minimal architectural state and memory required to replay the execution from the mid-point of the model we started recording from. The replayer (\S\ref{subsec:design_replayer}) is a modified existing ISA-level simulator, which replays the execution trace and generates detailed raw \uarchal metrics. Lastly, the analyzer (\S\ref{subsec:design_analyzer}) takes the raw generated metrics and performs various analyses to pinpoint inefficiencies and provides a detailed view of the hardware utilization to the user. %

\name is entirely implementable in software, requiring only a step-debugger and a simulator, two pieces of software that are commonly co-developed with the accelerator (\cSW). Most importantly, \name can be used to analyze any production model that already used in production without requiring special recompilation (\cNOCOMP). We have incorporated \name in our performance analysis workflow, which we describe in \S\ref{design:workflow}.

The rest of this section will focus on the design and implementation of the \name system, while the next section (\S\ref{sec:evaluation}) will dive deeper into the analyses developed with \name and how they answer the questions we posed.

\subsection{Execution Recorder}
\label{subsec:design_recorder}

The first step for \name is to capture the code running on the accelerator, along with any architectural state required to replay that code.
It is necessary to record the execution trace directly from the accelerator, rather than taking the compiler output, as the code can contain conditional executions and loops, and the contents of memory and registers are unknown at compile time.

An alternative to instruction traces might be to capture targeted performance traces using an instrumentation engine.
However, this would require recompilation (\cNOCOMP), which we want to avoid as it is both challenging in our environment, and we want to profile our models with their production compilation settings. Instead, we take advantage of a step debugger to capture these traces.

\textbf{Step debuggers.}
Step debuggers are a common software tool that is typically developed along with any accelerator. 
A step debugger uses hardware support to set breakpoints in the machine-code, which will pause the execution once the breakpoint address is hit. 
Users can then explore the contents of the memory, registers, or execute the next instructions one at a time (\textit{single stepping}). 
Because it is such a fundamental tool, all major vendors that we are aware of offer a step debugger for their ML accelerators (\eg NVidia CUDA-gdb~\cite{nvidia-cuda-gdb}, Cerebras CSDB~\cite{cerebras-step-debugger}).

We require the step debugger to provide three simple functions: a \texttt{step} function to execute instructions one-by-one, \asaf{I'm a bit confused about this -- how does the step debugger handle the fact that instructions are executed in parallel across many threads/warps? How would they be ordered?} \yannis{TPUs basically have 1 thread of execution. Each cycle, it executes ~10 instructions in-parallel (called a bundle). The parallelism is hardcoded in the hardware. For example, a matmul on a GPU would be spread across Streaming Processors. In the TPU, there's a big parallel MXU unit that implements matmul at the hardware level.} \texttt{read\_memory} and \texttt{read\_register} functions to read memory and registers respectively. 
To use the recorder, the user needs to specify a breakpoint at a location of interest and how many instructions to record. 
Once the breakpoint is hit, the recorder uses the \texttt{step} functionality of the debugger to execute and record instructions one-by-one.  A sketch of the recorder's logic is shown in Algorithm~\ref{algo:recorder}.  We'll now briefly describe the intuition behind it.

Each instruction takes zero or more inputs from input registers, and can write to zero or more output registers, or modify a memory region on the accelerator (\eg in the case of a DMA). \asaf{Maybe point to specific line numbers in the pseudocode}
In addition, some instructions (like DMAs) read directly from  device memory.
So, to accurately replay an instruction, we need to know the contents of these input registers and memory regions. 
To naively capture this information, we must first record the entire contents of all memory regions on the accelerator, as well as all registers that can be used as inputs to instructions. However, this can make the trace file very large. 
As an optimization, we recognize that we only need to store the contents of a register or memory region if those contents are \textit{first} used as input (and not as output) by a traced instruction. This is important to avoid recording intermediate results, which will be reconstructed in the simulator and are unnecessary to capture. \asaf{Maybe explicitly say here this is because you actually rerun the trace within the simulator} To avoid capturing this unnecessary information, the recorder maintains a set of each instruction's output registers and memory region addresses, and then only saves each instruction's input register contents or input memory region contents if those contents had not been modified by a previous instruction.

\begin{algorithm}
\caption{Execution Recorder Algorithm}
\label{algo:recorder}
\algsetup{linenodelimiter=\ }
\begin{algorithmic}[1]
\STATE $R \gets \emptyset$ \COMMENT{Set of used registers} 
\STATE $M \gets \emptyset$ \COMMENT{Set of used memory regions}
\FOR{$instruction\_count = 1$ \TO $N$}
    \STATE Parse instruction input register IDs and memory region addresses.
    \STATE Save READ\_REGISTER($r_i$); $\forall$ input register $r_i \notin R$
    \STATE Save READ\_MEMORY($m_i$); $\forall $ input memory region $ m_i \notin M$
    \STATE Parse instruction output register IDs and memory region addresses.
    \STATE $R \gets R \cup r_o; \forall r_o \in$ Output register IDs
    \STATE $M \gets M \cup m_o; \forall m_o \in$ Output memory regions

    \STATE Step instruction.
\ENDFOR
\end{algorithmic}
\end{algorithm}

\subsection{Replayer}
\label{subsec:design_replayer}

The second step in the \name methodology is to replay the captured trace of the model in a simulator and capture detailed metrics about the underlying \uarch of the accelerator. 
This is realized by the \emph{replayer}, a component based on an existing ISA-level simulator for the accelerator.

\name's key insight is to reuse the already available hardware simulator of the accelerator for the purpose of performance analysis. 
Hardware simulators, often called Golden Reference Models (GRMs) \asaf{I think you've already introduced these models several times earlier in the paper. Also you could consider moving this paragraph into the background section, and having a small mini subsection providing some background on GRM, it seems like it makes more sense to put this text there rather than in the design section}, are an artifact of the integrated circuit design process, where they help validate hardware design decisions before the final production~\cite{grm_ref_1,grm_ref_2,grm_ref_3}. 
These simulators implement the full instruction set in software, accurately modeling the architectural state inside the accelerator. 
Prior work has used simulators to try and evaluate Nvidia's GPU performance~\cite{gpgpu_sim, cuda_sim, ml_sim, accelsim}. However, due to the proprietary nature of GPU architectures, these studies could only approximate the actual design and exhibit significant errors. In our case, as both the vendor and consumer of our in-house accelerator, we have access to a fully accurate simulator and can provide a new perspective in using it for performance analysis.

The \name replayer repurposes this artifact of the design process for performance analysis, by extending its capabilities to capture metrics. 
We add a component to the GRM simulator, the \emph{performance tracker} which tracks metrics of interest. 
The performance tracker is passed as a dependency of the simulation, and registers callback functions with the modelled architectural components. 
On performance events, such as a memory read, the GRM executes the callback to the performance tracker, which records the event. 
When loading a trace, we first modify the memory and register state in the simulator to match the starting condition in our trace, then resume the execution of the recorded instructions. 
The performance tracker collects performance event information as the trace runs, which is finally saved to a metrics file for further analysis.

\subsection{Analyzer}
\label{subsec:design_analyzer}

The third step in the \name methodology is to analyze the captured events and metrics from the replayer. 
Our analysis framework is extensible and can examine any \uarchal component which is modeled in our GRM Simulator.
We briefly describe three sample analyses here, with full detail and results in our evaluation (\S\ref{sec:evaluation}).

The focus of our first analysis is the DMA subsystem.
Memory bandwidth is one of the most precious accelerator resources and for this reason we want to understand how the model interacts with system memory at a fine-grained level (\qMEM). 
Using the replayer's data, we construct a timeline view of all memory transfers along with their stalls. 
Users can leverage the information from this analysis to pinpoint inefficient DMAs. 
This analysis also reveals optimization opportunities across HLOs, as it is often hard to hide the DMA latency within a single HLO, but possible by looking further (\qXHLO).

Our second analysis focuses on analyzing the fine grained compute and memory utilization.
Providing instruction-by-instruction utilization information can help developers reason about resource availability at every step of execution (\qMXU) and debug low utilization of the hardware.

Finally we add an extension to \name to track instruction dependencies of the \assembly code in each trace.
\name can track all accesses to registers and memory locations in the simulator, capturing the data dependencies across instructions.
We can then use this dependency information to automatically suggest which instructions can be issued earlier (\qACT). We elaborate on how these analyses can be combined and provide valuable insights in \S\ref{sec:evaluation}.

\subsection{Workflow}
\label{design:workflow}

Finally, we describe where \name fits in the performance analysis workflow. This workflow is a result of our experience using \name to analyze the performance of several ML models. It consists of 1) classic \textbf{profiling} to gather regions of interest, 2) \textbf{recording} a trace, 3) \textbf{replaying} the trace in the simulator, and lastly, 4) \textbf{analyzing and visualizing}. 
We first collect some initial profiling with our in house performance profiling tool, which has similar functionality to Tensorboard~\cite{tensorboard}. This gives us a coarse-grained overview of performance and where potential bottlenecks might be. We note the instruction addresses of any areas of interest to record with our execution recorder (\S\ref{subsec:design_recorder}). To record the execution trace, we launch the execution recorder on the accelerator host machine, which attaches to a local accelerator running the model and sets a breakpoint for the target instruction address. Once the breakpoint is hit, the recorder takes over and creates execution traces, which are then saved remotely. These traces are typically 100-600k instructions long, which corresponds to a couple of model layers. These traces then serve as the input to the replayer (\S\ref{subsec:design_replayer}), which replays the trace and outputs performance metrics. Finally, the raw metrics are fed into the analyzer to perform the various performance analyses.

Note that so far we have been recording models in a sandbox environment, as setting a breakpoint in a customer-facing model would introduce unacceptable latency. However there is no technical restriction that keeps us from attaching the recorder to an ML model running in production.

\section{Evaluating Models With \name}
\label{sec:evaluation}

Armed with \name, we can now find out the answers to our research questions (\S\ref{sec:design_requirements}). First, \name's DMA analysis (\S\ref{subsubsec:dma_stall_analysis}) reveals that our models frequently incur unnecessary stalls waiting for memory transfers to complete, because of ineffective DMA scheduling. Second, \name's \uarchal utilization analysis helps users debug issues that aggregate metrics overlook, such as identifying the root cause of low compute utilization (\S\ref{sec:evaluation:compute_util}). Finally, \name enables the user to reason about possible optimizations, by providing detailed information about the \datacache utilization (\S\ref{sec:vmem_util}) and the dependencies of each instruction (\S\ref{subsubsec:dep_tracking}). By assembling a complete view of the system, \name is even able to automatically suggest optimizations via instruction re-orderings.

\subsection{Experimental Setup}
We use \name to investigate inefficient operations on three of our most important in-house LLMs, shown in Table~\ref{tab:model_description}. Each model is compiled with our in-house compiler using production configurations and deployed in the same topology as on production machines.

\begin{table}[]
\begin{tabularx}{0.45\textwidth}{|l|X|X|l|}
\hline
\textbf{Model} & \textbf{\#Params} & \textbf{\#Accel} & \textbf{Speedup} \\ \hline
\modelone  & <10b & N & 1.0\% \\ \hline
\modeltwo & 10b-100b & 4N & 4.1\% \\ \hline
\modelthree & >100b & 16N & 0.14\% \\ \hline
\end{tabularx}
\caption{Model descriptions. We evaluate three important in-house LLMs.}
\label{tab:model_description}
\end{table}

\amanda{config discussion?}
\amanda{8b is single chip i think, 26b is two, and then idk how many are in 141b}

\subsection{DMA Stall Analysis} 
\label{subsubsec:dma_stall_analysis}

We illustrate \name's effectiveness by describing a set of optimizations found using \name within the DMA subsystem of our accelerator.
Memory size and bandwidth are critical for modern ML accelerators, especially as larger models become increasingly memory-bound. To maximize memory utilization, we use \name to analyze DMA performance and uncover inefficiencies (\qXHLO, \qMEM).

\begin{figure}[h]
    \centering
    \includegraphics[width=0.40\textwidth]{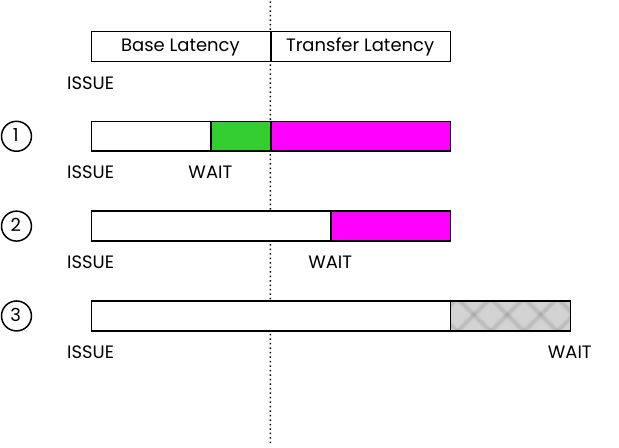}
    \caption{Lifetime of a DMA. A DMA is comprised of a base latency (constant) and a transfer latency (variable). DMAs begin with the ISSUE command. If the DMA has not completed by the time the WAIT command executes, the accelerator will stall until the DMA completes. We highlight three distinct scenarios. (1) WAIT comes before the base latency is fulfilled. The DMA incurs stalls first because of the base latency (green) and then because of the transfer latency (purple). (2) WAIT comes after the base latency but before the transfer latency is fulfilled. The DMA incurs stalls only because of the transfer latency (purple). (3) WAIT comes after the DMA finishes. The time between the DMA completion and the WAIT is slack (gray cross pattern).}
    \label{fig:dma-stall-types}
\end{figure}

\paragraph{Background on DMAs}
DMAs are a mechanism used by ML accelerators to transfer data between high-bandwidth memory (HBM) and the various caches. 
In our accelerator, DMAs are performed by an asynchronous DMA engine. 
They are controlled with two simple \assembly instructions as shown in Figure~\ref{fig:dma_assembly}. \dmaissue will start a memory transfer and \dmawait will block execution (sometimes stalling) until the transfer completes. 
Figure~\ref{fig:dma-stall-types} shows the lifetime of a DMA. Once issued, a delay occurs before the command reaches the DMA engine for processing and queuing. This delay, called base latency ($T_b$), is constant and non-accumulative: when multiple DMAs are issued in parallel, their base latencies are fulfilled simultaneously. Once the base latency is fulfilled, the DMA waits until it reaches the start of the engine queue and starts transferring data between the various memories. This delay is called the transfer latency, or $T_t$, and depends on the available bandwidth of the memories involved, the contention with other DMAs and the size of the transfer. Data transfer on a single link (source-destination pair) cannot be parallelized. Given the above, we can now examine the three scenarios that can happen for a DMA, as shown in Figure~\ref{fig:dma-stall-types}. The first case is when the WAIT instruction comes before the base latency is fulfilled. In that case, the DMA incurs stalls first because of the base latency, pictured as green, and then because of the transfer latency, pictured as purple. The second case is when the WAIT instruction comes after the base latency is fulfilled and the incurred stalls are only because of the transfer latency. The third case is when the WAIT comes after the DMA has completed. In that case, we say the DMA has "slack", as the \dmawait can be moved earlier. Slack is denoted in gray with a cross pattern.

\begin{figure}[h!]
    \centering
    \begin{tcolorbox}[colframe=black, colback=white, width=0.40\textwidth, title=\centering \textbf{ML Model Machine Code}]
        \begin{flushleft}
            \textbf{ISSUE} $\langle$dma settings$\rangle$ \\
            $\langle$other instructions$\rangle$ \\
            \textbf{WAIT} $\langle$dma\_id$\rangle$ \\
            $\langle$instruction using transferred memory$\rangle$
        \end{flushleft}
    \end{tcolorbox}
    \caption{Example \assembly instructions for handling memory transfers in an ML model. The ISSUE command starts the DMA, while the WAIT command blocks until it is complete and is typically inserted close to the command that needs the memory. The compiler can hide the DMA latency by inserting instructions between ISSUE and WAIT.}
    \label{fig:dma_assembly}
\end{figure}

The DMA subsystem is a common source of stalls because \dmaissue and \dmawait instructions can be difficult to properly schedule. In order to avoid incurring stalls, the compiler must insert enough instructions between each ISSUE and WAIT, so that it can hide the DMA's latency. This could be very difficult or impossible, especially within the boundaries of a single HLO.
Transfer stalls are also difficult to predict since transfers may be predicated, exist across multiple chips, and share limited bandwidth resources.
We differentiate between these two types of stalls in our analysis to help end users understand and remedy each scenario.

To construct the DMA analysis, \name captures DMA \dmaissue and DMA \dmawait in the replayer as it replays the trace. 
Then, given these events and the specifications about the transfer speeds of various memories, \name can simulate these transfers and flag which ones will stall the program. 
\name then plots the transfers in a timeline plot, highlighting the stalled parts.
The goal of this analysis is to visualize DMAs in an intuitive way and clearly show areas of potential improvements.

\subsubsection{Collective Operations Optimization}
\label{sec:all-gather}

Collective operations (like All-Gather, All-To-All, \etc) are fundamental operations for distributing a model across accelerators~\cite{openxla,nvidia-collectives-lib, aws-neuron-collectives-hw-support}. 
In this section, we describe our experience using \name to profile and optimize an important communication collective, \AllGather, reducing its runtime by 15\%.

We began our analysis by using our existing in-house start-of-the-art profiling tool to look at one of our most used models (\modelone). We observed that during generation for this model, the \AllGather operation was 13.3\% of total runtime. Our existing tool also reported that roughly 40\% of \AllGather execution time was spent stalling on DMAs. Beyond this information, we did not know what memory channels these stalls were coming from, what data was being transferred, or whether the stalls could be eliminated.

\begin{figure}[ht]
    \centering
    \draftfigure{\includegraphics[width=0.45\textwidth]{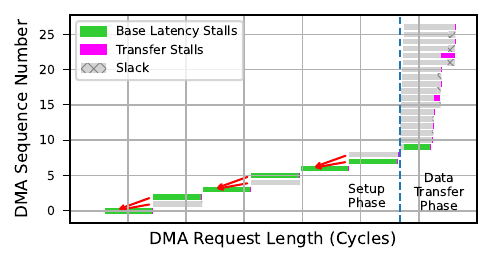}}
    \caption{All-Gather DMA Pattern. Memory accesses are performed in two phases, the setup phase, and data transfer phase. Dependencies for the setup phase
    are shown with red arrows. These dependencies were manually discovered by reading the \assembly, a laborious process.}
    \label{fig:all-gather-dma}

\end{figure}

Using \name, we capture a trace of 100,000 instructions for \modelone, and replay it on our modified ISA-simulator. The DMA stall analysis in Figure~\ref{fig:all-gather-dma} shows most stalls happen in the first nine DMAs. These stalls are mainly due to base latency (shown in green), not transfer latency (shown in purple), as the data transferred is small. The transfers go from HBM to the \datacache. We identify two optimization strategies. 
The first strategy is to issue these DMAs in parallel, if possible, since base latencies can be parallelized in our DMA engine. 
Manual dependency analysis of the \assembly shows the DMAs form three groups. 
Each group has an initial DMA, followed by two dependent DMAs. 
Issuing the three initial DMAs in parallel, then the remaining six in parallel, would reduce stalls by a factor of three. 
The second strategy requires more knowledge about how the All-Gather is implemented. 
The first nine DMAs are part of the operation setup phase, in which the addresses of other nodes in the topology are loaded. 
These addresses are then used as destinations for the DMAs in the subsequent data transfer phase. 
Since these addresses typically use a small amount of memory, especially with smaller models, our insight is that they could be permanently pinned in the \datacache. 
This would completely remove the need for the setup phase DMAs, eliminating the stalls they incur. 
However in larger models, where topology sizes are larger, pinning the node addresses in memory may consume too much \datacache memory. 

\begin{figure}[ht]
    \centering
    \includegraphics[width=0.40\textwidth]{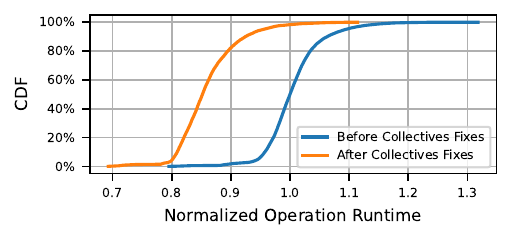}
    \caption{Reduction in runtime for the \AllGather collective operation after eliminating unnecessary DMAs. Runtime is normalized to the median value before optimization. }
    \label{fig:all-gather-improvement}

\end{figure}

After concluding our analysis, we worked with internal developers to optimize the collective. 
Our internal compiler team opted to implement the second approach (pinning the node addresses in memory), because of its lower implementation complexity. 
The optimized code reduced the runtime of the All-Gather operation by 15\%, as shown in Figure~\ref{fig:all-gather-improvement}. 
In terms of end-to-end runtime, it improved generation latency by 4\% in \modeltwo and 1\% in \modelone, as shown in Table~\ref{tab:model_description}. 
\modelthree's improvement was a smaller 0.14\%, highlighting the trade-off of the second approach. We are in the process of implementing our first approach (parallelizing setup phase DMAs), which is expected to perform better on larger models.

In conclusion, we used \name to diagnose and optimize several important collectives operations in our in-house LLMs. Our existing state-of-the-art internal tools could only tell us that these operations were stalling, but could not provide the actionable insights that were obvious with \name. Because these collectives operations are so crucial to LLM performance, our optimization lead to a realized improvement in end-to-end performance for our models. Every percent improvement in model serving leads to significant savings in total cost \asaf{could we say something here like tens of millions, hundreds of millions or so?} to run our models.

\subsection{Detailed \Uarchal Utilization}
\label{sec:vmem_util}
\amanda{\qMEM \qMXU}

ML accelerators include an assortment of compute units (\eg matrix unit, vector unit, scalar unit) and memories (\eg HBM, \datacache). Underutilization in any of these units is an opportunity for further optimizations. Existing tools like Google's Tensorboard provide coarse-grained metrics about higher level functions. While an aggregated utilization is useful as a measure of performance, it does not give developers a a full picture of how the accelerator is used or, more importantly, what could be changed. In this section, we show how \name's detailed hardware utilization analysis enabled us to improve the runtime of a compute-heavy operation by 70\%.

\subsubsection{Compute Unit Utilization}
\label{sec:evaluation:compute_util}

\begin{figure}[h!]
    \begin{subfigure}[c]{0.40\textwidth}
        \centering
        \includegraphics[width=\textwidth]{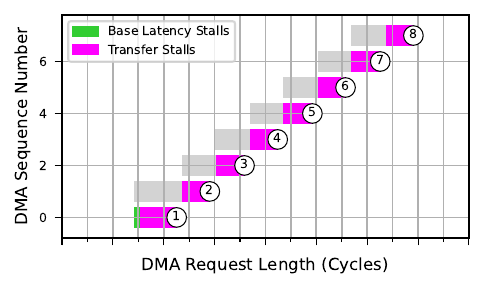}
        \caption{DMA Sequence. \amanda{put HLO lines}}
        \label{fig:dep-tracking:dmas}
    \end{subfigure} \\
    \begin{subfigure}[c]{0.40\textwidth}
        \centering
        \includegraphics[width=\textwidth]{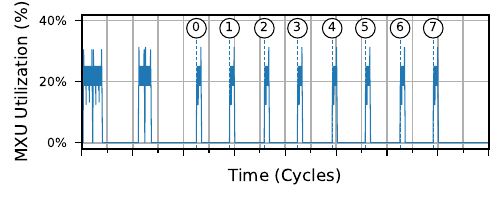}
        \caption{Matrix Multiplication Utilization}
        \label{fig:dep-tracking:mxu}
    \end{subfigure} \\
    \begin{subfigure}[c]{0.40\textwidth}
        \centering
        \includegraphics[width=\textwidth]{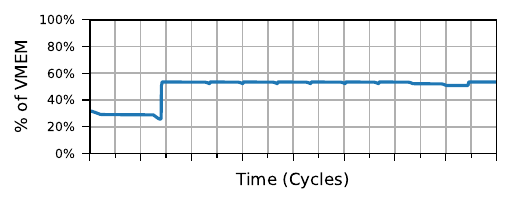}
        \caption{\Datacache Utilization}
        \label{fig:dep-tracking:vmem}
    \end{subfigure}
    \caption{\name can make it easy to diagnose low matrix multiplication utilization. We annotate the completion of each DMA, and the corresponding cycle on the \MXU utilization (Figure~\ref{fig:dep-tracking:mxu}).  We see that the matrix multiplier is waiting on data transfers, resulting in low utilization. At the same time, the \datacache has enough is not fully utilized, so we might be able to prefetch more data.
    }
    \label{fig:dep-tracking}

\end{figure}

\begin{figure*}[h!]
    \centering
     \begin{subfigure}[b]{0.75\textwidth}
         \centering
         \includegraphics[width=\textwidth]{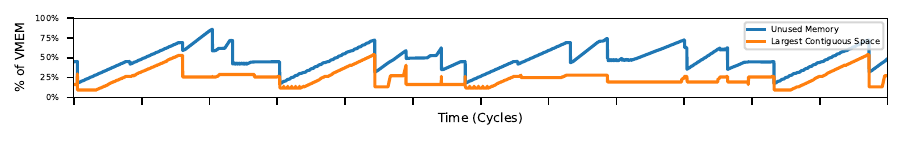}
         \caption{Total Occupancy}
         \label{fig:vmem-fragmentation:util}
     \end{subfigure}
    
    \begin{subfigure}[b]{0.75\textwidth}
        \centering
        \includegraphics[width=\textwidth]{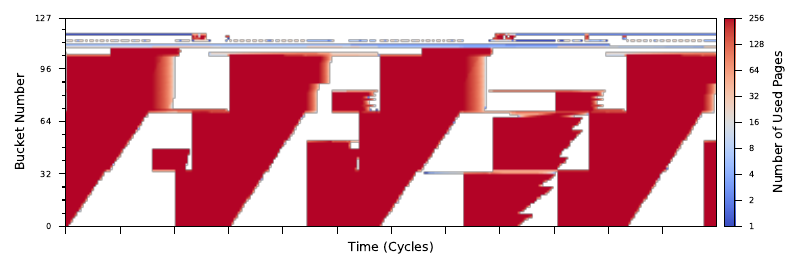}
        \caption{Fragmentation Heatmap}
        \label{fig:vmem-fragmentation:heatmap}
    \end{subfigure}
    \caption{Figure~\ref{fig:vmem-fragmentation:util} shows the percentage of \datacache space which is unused, and the percentage which is unused and contiguous. The largest contiguous region is consistently much smaller than the total unused portion. Figure~\ref{fig:vmem-fragmentation:heatmap} is the \datacache fragmentation analysis. The total memory space is divided into pages, where each read and write is tracked. Pages are then grouped into ``buckets'' of 256 pages. Each bucket is a horizontal line in the plot and its color denotes the number of used pages in the bucket for each cycle. We see considerable fragmentation of our \datacache, which is critical for performance.  }
    \label{fig:vmem-fragmentation}

\end{figure*}

To aid users in maximizing utilization of the compute units in an accelerator, we would like to provide them with a view of how the model interacts with the hardware at very high detail. Thanks to its use of a hardware simulator, \name can reveal how well the model utilizes the compute units at single cycle granularity. To understand how this can assist users, we provide an example of debugging low utilization of the matrix multiplication unit (\MXU) in a real model\footnote{Examples of matrix multiplication units include the MXU for Google TPUs and the TensorCore for Nvidia GPUs.}.

Figure~\ref{fig:dep-tracking} shows \name's analysis for a model section that exhibits low compute utilization. Our high level tool simply reports a low compute unit utilization, making it hard to dig deeper into the root cause. In contrast, \name's analysis directly points out the problem. First, Figure~\ref{fig:dep-tracking:mxu} shows the utilization of the matrix multiplication unit, the main workhorse of the accelerator. 
We see an intermittent pattern of computation and data transfer, indicating that the problem is caused by some other component bottlenecking the flow to the \MXU.  
We can immediately spot the problem by using \name to look at the memory subsystem (Figure~\ref{fig:dep-tracking:dmas}). The high utilization coincide with DMA completions and the zero utilization spots coincide with the DMA stalls. The current DMA schedule is not feeding data to the matrix unit fast enough. \asaf{Maybe explain more clearly what the labels (the numbers) in the figure denote}

We see how an issue that was opaque before now becomes clear to the user, who can reason about optimizing it. A straightforward path from this finding would be to prefetch more aggressively before the chain of matrix operations begins. However, this requires reasoning about the availability of \datacache at that point in time. 
Because \name also allows us to track this information, we can investigate \datacache utilization (plotted in Figure~\ref{fig:dep-tracking:vmem}), and we see that there is sufficient room in our \datacache to prefetch our data. Finally, by manually analyzing the \assembly code, we conclude that these DMAs dependencies allow them to be issued earlier.
With this information, we consulted with our internal compiler team to implement the prefetching, lowering operation runtime by 70\% and increasing \MXU utilization.

\subsubsection{Memory and Data-Cache Utilization}
\label{sec:evaluation:vmem_fragmentation}

In our previous example, we showed how examining the overall utilization of the \datacache can help users make decisions about data prefetching. 
However, \name can analyze the \datacache at a much finer grained level, enabling developers to not only look at overall utilization but also cycle by cycle memory usage and fragmentation.

\paragraph{Background on the \datacache} 
The \datacache is a relatively small, fast cache that stores vectors that can quickly be loaded into the matrix multiplication units. 
The \datacache bridges the slow HBM with the fast compute capability of the accelerator, making it crucial to model performance. 
The \datacache and memory allocations are static and managed by the compiler during the compilation process. However, because of predication and loops, it is hard to reason about the exact \datacache use at each point of the program. Moreover, a challenge with these memory allocation techniques is fragmentation. It is not enough to have a large on-chip memory, there also must be a contiguous region available as a target for memory transfer.

By using a hardware simulator, \name can precisely track usage of the \datacache by tracking reads and writes on \datacache pages.
A page is counted as ``used'' if it is read by an instruction after being written.
If a page is brought in from HBM to the \datacache and the page is not read in subsequent instructions then that page is tracked as ``unused''. We are only able to track these categories as long as our simulation runs, so theoretically if a page is read after the simulation ends, our analysis would miss that read. However, our simulations typically run for hundreds of thousands of cycles. A memory transfer should be read within that period, and if a page is transferred but not used for tens or hundreds of thousands of cycles, that is an inefficiency in the model.

The \datacache fragmentation analysis is shown for a production LLM in Figure~\ref{fig:vmem-fragmentation}. 
The total \datacache space is divided into 128 ``blocks'', each of which contain 256 pages.
The heatmap information shows how many pages were used within each block. 
In Figure~\ref{fig:vmem-fragmentation:util}, we show both the percentage of unused memory, and the percentage of memory occupied by the largest contiguous region. 
The total portion of unused memory varies between approximately 20\% and 80\%, however the largest contiguous unused space is consistently less than this. 
In this trace, the median unused space is 47\% of the total space, and the median contiguous block is only 25\% of the total space. 
These results show that the \datacache space can be used more efficiently. 
\asaf{It's not clear though what's the actionable insight here}

Beyond model developers, this analysis provides compiler engineers with a unique insight into \datacache utilization at the lowest level. Memory allocation is done by the compiler and relies on finely-tuned heuristics.
 Compiler engineers often evaluate memory allocation techniques on models that quickly become outdated and don't have a way to gain insight into the way these heuristics are affecting the latest production models. \name offers a new way of debugging these complex compiler subsystems that wasn't possible before.

\subsection{Dependency Analysis}
\label{subsubsec:dep_tracking}
\amanda{\qXHLO, \qMXU, \qACT}

In the previous examples, we demonstrated how \name's fine-grained analysis can identify inefficiencies. However, it's difficult to differentiate between poor performance which is actually unavoidable due to dependencies, and poor performance which can be eliminated through alternative schedules. So far, we tackled the issue by manually going through the \assembly and discovering dependencies, which is a labor intensive and error prone process. Automating the discovery of these dependencies would not only alleviate the need for users to manually analyze complex \assembly code but also enable \name to autonomously propose optimizations through alternative instruction scheduling. In this section, we present how \name implements dependency tracking and leverages this information to enhance the effectiveness of our existing analyses.

\yannis{There's a subtle difference between dep analysis and program slicing. The first focuses on instruction deps for optimization, the second appears more in software engineering literature and attempts to identify the pieces of source code affecting a computation. Because of this, I'm a bit hesitant to mention it.}

There are two ways to do dependency analysis: static and dynamic. 
While static analysis can be done by the compiler, a static implementation will be missing key information about runtime behavior, such as predication, which limits its usefulness. 
\name implements dynamic dependency analysis, using the GRM simulator to discover instruction dependencies as it replays the trace. 
Instruction dependencies can be registers (\eg for arithmetic instructions), \datacache locations (\eg for load/store instructions) or even HBM locations (\eg for DMA instructions). 
To discover instruction dependencies, \name traces all accesses to registers, \datacache and HBM in the simulator. 
When an instruction reads from a register or piece of memory that was previously written by another instruction, then those instructions have a dependency. 
With this information, \name knows the exact dependency graph between instructions.

\begin{figure}[h]
\centering
\includegraphics[width=0.30\textwidth]{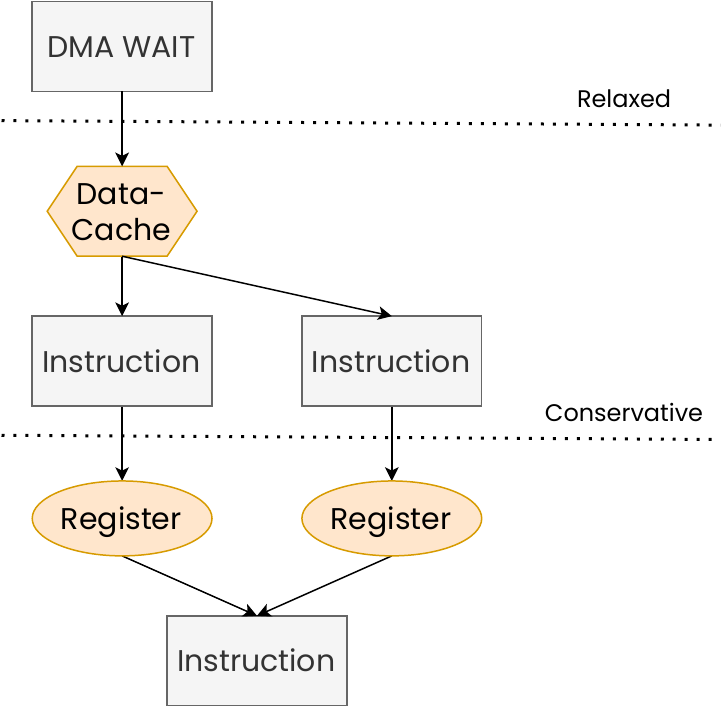}

\caption{Different models of dependency tracking. In the conservative models, an instruction depends on the immediate instructions that shape its input. In the relaxed model, dependencies are propagated until the dependent memory is in the \datacache.}
\label{fig:dep-tracking-models}

\end{figure}

One use case for this analysis is to reason about rescheduling DMAs. 
The way DMAs are typically issued is to load some information about the DMA metadata from HBM. 
This metadata is put in registers and may be transformed before being used as input to the DMA instruction. 
These transformations are typically lightweight and they occur close to the DMA issue instruction. 
If we only track when a DMA instruction's immediate input registers as dependencies, then in many cases the dependencies will be flagged as ``fulfilled'' immediately preceding the DMA. 
However, our insight is that as long as any transformations on the DMA metadata are lightweight, we can move the DMA \dmaissue and any transformation instructions as early as when the data is ready in the \datacache. 
We call this method of dependency accounting ``relaxed'', illustrated in Figure~\ref{fig:dep-tracking-models}.
In the conservative model, an instruction depends on its input registers. 
In the relaxed model, an instruction depends on the DMA that brought its inputs, propagated through dependencies, to memory. For DMA instructions, it is better to use the relaxed model to reason about their dependencies.

\begin{figure}[h]
    \centering
    \includegraphics[width=0.40\textwidth]{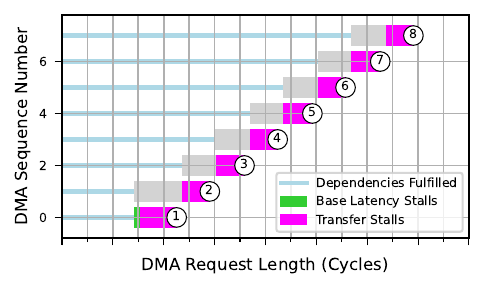}
    \caption{Example dependency analysis with \name, showing the DMA analysis of section ~\ref{sec:evaluation:compute_util} with dependency information overlayed as "backtails". Thanks to that, the user can immediately conclude the DMAs can be issued earlier, without needing to manually analyze the \assembly.}
    \label{fig:dep-tracking:example}
\end{figure}

Using the generated dependency graph, \name can display this information to the user, allowing them to reason about alternative schedules that eliminate stalls by reordering instructions. 
We illustrate the power of \name's dependency tracking by applying it to the example shown above, where we had to manually discover dependencies by reading \assembly code. 
Figure~\ref{fig:dep-tracking:example} shows the DMA analysis with dependency information overlayed. 
Dependencies are visualized as ``backtails'' for each DMA, indicating how far back we can start it based on the relaxed model. 
The user can see that all DMAs can be issued earlier in time and thus prefetching is a viable optimization strategy.

\paragraph{Autonomous Optimization Suggestions} 
By combining the various analyses presented so far, \name can reason about inefficient DMAs, check if alternative schedules are possible and flag them for further investigation. 
An outline of this logic is shown in Algorithm~\ref{algo:auto-suggestions}. 
We examine each DMA separately. 
If the DMA does not have stalls, we have nothing to suggest. 
Otherwise, we first check how far we can push the DMA based on its dependencies. 
Specifically, we would like to push it back at least as many instructions as stall cycles, to eliminate those stalls. 
Second, we check if the \datacache has enough contiguous memory available to accommodate the DMA, using the analysis from Section~\ref{sec:vmem_util}. 
If these two conditions are true, then we suggest a DMA reordering to eliminate stalls. This algorithm successfully discovers and suggests the optimizations from the two examples we showed.

\begin{algorithm}
\caption{Automatic optimization suggestion algorithm}
\label{algo:auto-suggestions}
\begin{algorithmic}[1]
\FOR{each DMA operation}
    \IF{not \textit{stalled}}
        \STATE \textbf{continue}
    \ENDIF
    \STATE $push\_limit \gets$ DMA dependency distance
    \STATE $memory\_available \gets$ (contiguous data-cache $\geq$ $dma$ size)
    \IF{push\_limit > stall\_duration \textbf{and} memory\_available}
        \STATE Suggest reordering
    \ENDIF
\ENDFOR
\end{algorithmic}
\end{algorithm}

In conclusion, dependency analysis enables \name to both aid the user in reasoning about what optimizations are possible, as well as automatically suggest them. \asaf{did you actually apply this approach to any real use case?}

\subsection{\name's Overhead}
\label{sec:evaluation:overhead}

Since \name traces every instruction in an ML model and replays it in a \uarchal simulator, the overhead is high. Thus, a key concern is whether \name captures enough of the model’s execution or zooms in too narrowly, missing context.

Currently, \name records ~400 us of model execution, covering multiple layers. The snapshot size is about 100MB. The end-to-end process takes around 2 minutes, but can be easily optimized to under a minute. Given the model's self-similarity, capturing a few layers across different modes (\ie prefill, decode) provides sufficient performance insights. Overall, \name's overhead is manageable, with potential for further optimization.

\section{Conclusion}

The exploding demand for ML accelerators mandates squeezing the most out of the existing infrastructure. \name is a novel approach for analyzing performance by bridging the gap between high level semantics,  machine code, and micro-architecture - revealing new opportunities in an already highly-optimized environment. 
Other hyperscalers can reproduce \name approach on their custom accelerators, given fairly basic prerequisites: a step debugger and an ISA simulator. The yields of optimizing large ML workloads will enable developing more sophisticated models while significantly reducing both capital and power requirements.

\section{Acknowledgements}

We thank the TPU compiler team, the Gemini team, and all the teams working across the ML stack at Google for their continuous support throughout this work. We also thank Kostis Kaffes for his insightful guidance and comments in writing this paper.

Ioannis Zarkadas is an Onassis Foundation scholar.

\bibliographystyle{ACM-Reference-Format}

\bibliography{refs}

\end{document}